\documentstyle[prl,aps,twocolumn,epsfig]{revtex}

\begin{document}
\date{\today}
\title{\bf{Collective Dipole Bremsstrahlung in Fusion Reactions}}

\author{V. Baran$^{ab)}$, D.M. Brink$^c$, M. Colonna$^{a)}$, M. Di Toro$^{a)}$}

\address{$^{a)}$ Laboratori Nazionali del Sud, Via S. Sofia 44,
I-95123 Catania, Italy and University of Catania, \\
$^{b)}$ NIPNE-HH, Bucharest, Romania,
$^{c)}$ Oxford University, U.K.}

\maketitle

\begin{abstract}
We estimate the dipole radiation emitted in fusion processes. We show 
that a classical bremsstrahlung approach can account for both the preequilibrium
and the thermal photon emission. We give an absolute evaluation of the
pre-equilibrium component due to the charge asymmetry in the entrance 
channel and we study the energy and mass dependence in order to optimize
the observation. This dynamical dipole radiation could be a relevant cooling
mechanism in the fusion path. We stress the interest in experiments with the
new available radioactive beams. 
\end{abstract}

\hspace{-\parindent} PACS numbers: 24.30.Cz; 25.70.Jj; 25.70.Lm; 25.60.Pj

In the early stage of a heavy ion collision large amplitude collective
motions can be excited. Monopole, dipole or quadrupole modes are typical
examples. Tracing information from their decay provides the opportunity to
learn about the features of fermionic systems in particular, extreme
conditions.  

If the colliding nuclei have different $N/Z$ ratios the charge equilibration
process takes place. Several works have suggested that the related
neutron-proton collective motion has the features of a Giant
Dipole Resonance ($GDR$) mode
\cite{hof79,bon81,ditoro,sur89,cho93,bar96,ced01}. In a microscopic
approach based on semiclassical $BNV$ transport
equations \cite{bar01}, it has been studied in detail how a collective $GDR$ 
responsenbdevelops in the entrance channel. In connection to this analysis
three main phases can be identified.
First, during the ${\it 
approaching~phase}$ the two partners, overcoming the Coulomb barrier, still
keep their own response. Then it follows a {\it dinuclear phase}
when the conversion of relative motion energy in thermal motion starts to 
take place, mainly due to nucleon exchange.
The composite system is not thermally equilibrated
and  manifests, as a whole, a large amplitude  dipole collective motion.
The charge equilibration has initiated.
Finally, during the third phase, the {\it Compound Nucleus (CN) formation},
the nucleus is thermally equilibrated.
In a phonon approach \cite{brink90,bor91} it was shown \cite{cho93} for
the first time that an enhancement of the $GDR$ gamma emission should
be observed if the number of $GDR$ phonons, $n_{GDR}^{(0)}$,
when the $CN$ is formed, is
larger than the value corresponding to a statistical equilibrium
of the $GDR$ (as oscillator) with the $CN$ (as heat bath). 
This effect has been experimentally evidenced in several fusion reactions
\cite{fli96,cin98,amo198,pie01}. 

However it is important to notice that also in the second {\it (dinuclear)}
phase a $GDR$ is excited on top of nonequilibrium states \cite{bri55}
(Brink-Axel hypothesis). The number of $GDR$ phonons is even larger than 
$n_{GDR}^{(0)}$ \cite{bar01} and consequently some  pre-equilibrium $GDR$
photons can be emitted also at this
stage. In the attempt to estimate this contribution we cannot use
straightforwardly the "phonon" approach \cite{cho93,brink90} which is based
on the assumption of the existence of a thermally equilibrated nucleus.
A way to face this problem is to directly apply a bremsstrahlung 
{\it ("bremss")}  approach.
Considering the evolution of the collective dipole acceleration from the time
when it suddenly rises, at the beginning of the second phase, until it is
completely damped to a pure "thermal" component, we can consistently 
calculate the whole contribution of the pre-equilibrium
$GDR$ to the photon yield. This is the aim of our letter.   

The total photon emission probability from the 
dipole mode oscillations, as given by the bremsstrahlung formula, can be
expressed as \cite{jak62} ($E_{\gamma}= \hbar \omega$):
\begin{equation}
\frac{dP}{dE_{\gamma}}=\frac{2 e^2}{3\pi \hbar c^3 E_{\gamma}}
(\frac{NZ}{A})^{2}
|X''(\omega)|^{2}  \label{brems},
\end{equation}
where $X''(\omega)$ is the Fourier transform of the acceleration $X''(t)$
associated with the  distance between the centers of mass of
protons $(P)$ and  neutrons $(N)$, $X=R_{p}-R_{n}$.
$A=N+Z$ is the system mass. 
Thus following the time evolution of the dipole mode along the fusion 
dynamics it is possible to evaluate, in absolute values, the corresponding
pre-equilibrium photon emission. 
However it is interesting to notice
that the {\it "bremss"}  
approach \cite{ber98} actually
provides a unified picture to account for the photon emission from both
pre-equilibrium and thermal $GDR$.  
In fact Eq.(1) also allows to recover the 
usual expression for the $\gamma$-decay rate from a $GDR$ in thermal 
equilibrium with a $CN$ at temperature $T$.
In this case the dipole mode 
will manifest thermal oscillations that 
by the fluctuation-dissipation theorem  
are completely determined {\it only} by its dissipative properties 
\cite{cal51}. We may consider equivalently that 
these are the consequence of the action of a fictious,
time dependent, random force $F(t)$ \cite{landau}.
Then the Fourier transform of the distance $X(t)$ can be expressed as: 
$X(\omega) = \alpha(\omega)\times F(\omega)$, where 
\begin{equation}
\alpha(\omega) = \alpha_r(\omega) + i \alpha_i(\omega) = 
\frac{1}{M_{coll} (\omega_{0}^2-\omega^{2}-
i  \frac{\Gamma}{\hbar} \omega)}, 
\end{equation}
is the response function of the dipole mode \cite{brink88}.   
$M_{coll} \equiv \frac{NZ}{A} m$ with $m=935MeV$
(nucleon mass) represents the collective mass of the
neutron-proton relative motion and $\Gamma$ is the decay width. 

The spectral density of mean square fluctuation of random force
$<|F(\omega)|^{2}>$ is related to the properties of the system  through the 
fluctuation-dissipation theorem, as it follows \cite{cal51,landau}: 
\begin{equation}
<|F(\omega)|^{2}>=\frac{\hbar \alpha_{i}(\omega)}{|\alpha(\omega)|^{2}}
cth(\frac{\hbar \omega}{2T})
= M_{coll} \omega \Gamma cth(\frac{\hbar \omega}{2T}). \label{fdt}
\end{equation}
Using the Parseval theorem and the ergodic hypothesis we have,
for times $T_{0}$ very large, 
$\lim_{T_{0} \rightarrow \infty} T_{0} <|F(\omega)|^{2}>
 = |F(\omega)|^{2}$. So, by introducing in Eq.(\ref{brems})
the Fourier transform $X''(\omega)$ of the
acceleration, since $|X''(\omega)|^{2} =\omega^4 |X(\omega)|^2$, we get
the average photon emission probability per unit time:
\begin{equation}
\lim_{T_{0} \rightarrow \infty} \frac{1}{T_{0}} \frac{dP}{dE_{\gamma}}=
\frac{2}{3\pi} \frac{e^2}{\hbar m c^{3}} \frac{NZ}{A}
\frac{\Gamma E_{\gamma}^{4}}{(E_{\gamma}^2-E_0^{2})^{2} +
\Gamma^2 E_{\gamma}^2}cth(\frac{\hbar \omega}{2T})
\end{equation}
If we exclude the zero-point motion contribution and expand the
hyperbolic function for $\hbar \omega > T$ we obtain: 
\begin{eqnarray}
\lim_{T_{0} \rightarrow \infty} \frac{1}{T_{0}} \frac{dP}{dE_{\gamma}} =  
\frac{1}{\hbar}(\frac{E_{\gamma}}{\pi \hbar c})^2~~~~~~~~~~~~~~~~~ 
\nonumber \\ 
\frac{1}{3} \frac{4 \pi e^2 \hbar}{mc} \frac{NZ}{A} 
\frac{\Gamma E_{\gamma}^{2}}{(E_{\gamma}^2-E_0^{2})^2 +
\Gamma^2 E_{\gamma}^2} \exp({-\frac{E_{\gamma}}{T}}) = \nonumber \\
\frac{1}{\hbar}(\frac{E_{\gamma}}{\pi \hbar c})^2
\frac{\sigma_{abs}}{3} \exp(-\frac{E_{\gamma}}{T}),~~~~~~~~~~~~~~~~
\label{statis}
\end{eqnarray} 
where $\sigma_{abs}$ is the $\gamma$-absorption cross section
in the $GDR$ region.
This is the well known formula for the gamma emission rate from hot $GDR$
\cite{sno86}.
It was previously obtained by using the detailed balance principle or
from considerations on the statistical equilibrium between dipole mode
and black-body radiation \cite{brink88}.

Now we focus on the dynamical dipole mode in 
the entrance channel and
we study the dependence of the pre-equilibrium dipole oscillations 
on the incident 
beam energy. We look at the relative weight vs. the statistical
contribution. 

In the following we present results concerning the charge equilibration
process for the systems $^{40}Ca$ $(N/Z=1)$ on  $^{100}Mo$ $(N/Z=1.38)$,
at $4AMeV$ and $^{16}O$ $(N/Z=1)$
on $^{98}Mo$ $(N/Z=1.33)$ at various beam energies
($4$, $8$, $14$ and $20AMeV$).
Our fusion dynamics is obtained in the framework of an accurate 
$BNV$ transport
simulation of central collisions, see \cite{bar01}
for a detailed analysis.

The time evolution of the dipole moment $D(t)=\frac{NZ}{A} X(t)$ 
and of quantity $DK(t)=\Pi/\hbar$  for the
$Ca+Mo$ reaction is shown in Fig.1a.
Here $\Pi= \frac{NZ}{A}(\frac{P_{p}}{Z}-\frac{P_{n}}{N})$
with $P_{p}$ ($P_{n}$) center of mass in momentum space for protons
(neutrons) is just the canonically conjugate momentum of the $X$ coordinate,
$\Pi=M_{coll} X'(t)$. 
Then a collective $GDR$ hamiltonian
can be introduced as
$H_{GDR} =\Pi^{2}/(2M_{coll}) + M_{coll} \omega_{0}^{2} X^{2}/2$.

We choose the origin of time at the beginning of the {\it dinuclear}
phase \cite{bar01}.
The dipole acceleration as obtained by a second order 
numerical derivative of $D(t)$ is shown in Fig.1b. With a good approximation
this is zero before $t=0$. It suddenly rises when the collective
dipole excitation is triggered and after few oscillations 
it becomes completely damped, around
$250 fm/c$. The corresponding power spectrum, $|D''(\omega)|^{2}$,
calculated from the Fourier transform of $D''(t)$:
\begin{equation}
D''(t)=\frac{1}{2\pi} \int_{0}^{\infty} D''(\omega)
\exp^{ -i \omega t} d \omega  \label{fourier}
\end{equation}
is plotted in Fig.1c by a dashed line.

\begin{figure}[htb]
\epsfysize=5.5cm
\centerline{\epsfbox{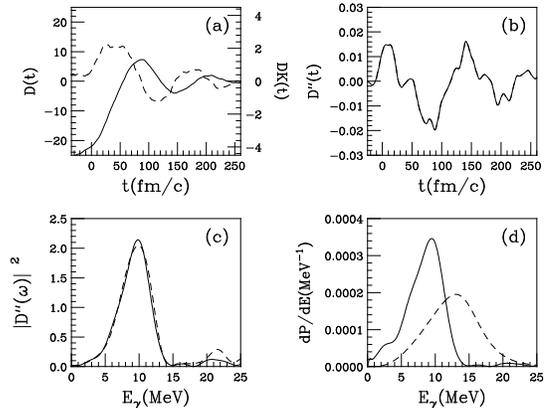}}
\caption{System $Ca$+$Mo$ at $4~AMeV$: (a)
Time evolution of $D(t)$ (solid line, in {\it fm units}),
and  $DK(t)$ (dashed line, in  $fm^{-1}$). (b) The same for the 
acceleration $D''(t)$
(in $c^{2}/fm$). (c) 
Power spectrum $|D''(\omega)|^{2}$ (in {\it $c^2$ units}).
 (d) Bremsstrahlung spectrum (solid line) and the first step statistical
spectrum (dashed line).}
\end{figure}
In the same Fig.1c we have added (solid line) $|D''(\omega)|^{2}$ as obtained
directly from the Fourier transform of the dipole moment, $D(\omega)$ by using:
\begin{eqnarray}
|D''(\omega)|^2 &=& D'(0)^2 +\omega^2 D(0)^2 + 2 \omega^2 D'(0) Re D(\omega) 
\nonumber \\
&-&2 \omega^3 D(0) Im D(\omega) + \omega^4 |D(\omega)|^2  \label{dipdip}
\end{eqnarray}
where $D(0)$ and $D'(0)$ are the dipole and the dipole velocity
values at $t=0$. The nice agreement between the two procedures
indicates the numerical accuracy of the method.

The total photon emission probability from the pre-equilibrium
dipole mode, given by the bremsstrahlung formula Eq.(\ref{brems}),
is reported in Fig.1d (solid line).

For comparison we show the first step statistical
spectrum (dashed line). The latter is just the 
product of the statistical gamma
decay rate \cite{sno86}, (see also Eq.(\ref{statis})), 
times the mean life time of the
compound nucleus $\tau_{CN}$. A good estimate of $\tau_{CN}$ can been 
derived
from the total neutron width of a $CN$ at temperature $T$:
$\Gamma_{n} = \hbar/\tau_{CN} =2m r_{0}^2 A^{2/3}/(\pi \hbar^2)
T^2 \exp(-B_{n}/T)$,
with $B_{n}=8.5 MeV$ (neutron binding energy) 
and $r_{0}=1.2 fm$.
For the statistical $GDR$, the standard parameters from systematics,
including temperature dependence of the width, are considered.
We remark that the two contributions are comparable. 

The pre-equilibrium spectrum is shifted toward lower values of energy as
a consequence of the large quadrupole deformation along the fusion path.
However part of the total shift is also due to
the used effective interaction \cite{bar96}.

We perform an analogous analysis for the reaction $^{16}O$ $+$ $^{98}Mo$
at four beam energies.
In Fig.2 the power spectra $|D''(\omega)|^{2}$  obtained 
with the two procedures described before
are plotted. Again the two methods give close results.

\begin{figure}[htb]
\epsfysize=5.5cm
\centerline{\epsfbox{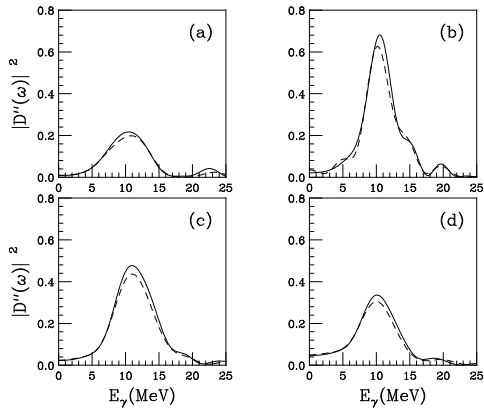}}
\caption{
Power spectrum $|D''(\omega)|^{2}$ for the  system $O$+$Mo$
at the beam energies $4$ (a), $8$ (b), $14$ (c) and $20$ (d)
$A MeV$ as
obtained directly from dipole acceleration (solid line) 
and from Eq.(\ref{dipdip}) (dashed line).  
}
\end{figure}

In Fig.3 the corresponding {\it "bremss"} spectra  
(solid lines) as well as the first step statistical spectra
(dashed line) for the system $O + Mo$ are 
plotted. We remark the characteristic {\it "rise and fall"}
 behaviour of the
pre-equilibrium contribution (already observed within the "phonon"
approach \cite{bar01}).
This effect is more clearly evidenced in Fig.4 
where the extra photon multiplicities due to the charge asymmetry
in the entrance channel,
obtained integrating over energy in the resonance region,  
are plotted as a function of excitation energy for both 
{\it "bremss"} and {\it "phonon"} (see following)
models.
The resulting trend is determined by the interplay
of several effects. At low beam energies, a slow neck dynamics does not
sustain the dipole oscillations. A strong attenuation of the acceleration  
takes place along entrance channel and the dipole emission is quite low.
We expect a similar behaviour in deep-inelastic reactions.

At higher beam energies a faster fusion dynamics can favor the dipole
oscillation. Larger amplitudes of the acceleration and more dipole 
oscillations enhance the pre-equilibrium radiation. Around 8-14 AMeV an
optimum effect is attained. At higher
beam energies a larger $GDR$ damping manifests and consequently
the gamma emission is again reduced \cite{bar01}.
 
From a comparison between the two systems, at the same energy
available in the c.m.,  we observe a larger
pre-equilibrium emission from $Ca+Mo$ (black square in Fig.4). 
 This can be
related to larger values of the dipole acceleration as a result
of a larger initial dipole moment amplitude.

It is interesting to compare our results to the predictions of
the {\it phonon} approach, where the main input quantity is the number of
$GDR$ phonons at the time of $CN$ formation. In this model a $GDR$ phonon
gas is coupled to the $CN$ \cite{cho93}. The phonons are decaying with a
rate $\mu=\Gamma/\hbar$ and excited with a rate $\lambda$. 
The enhancement of $\gamma$-decay probability is given by
$\Delta P_{phonon} = (n_{GDR}^{(0)}-\lambda/\mu)
\gamma_{\gamma}/(\mu+\gamma_{ev})$
were $\gamma_{\gamma}$ is the partial width for photon emission and
$\gamma_{ev}$ is the evaporation decay rate. In this method one of the
uncertainties is related to the estimation of the quantity $n_{GDR}^{(0)}$
which in turn requires the knowledge of the $CN$ formation time \cite{bar01}.
The results presented here are obtained following the usual
procedures in $\vec{r}-$ and $\vec{p}-$ space to fix the thermal 
equilibration time, see ref.\cite{bar01}.
We see from Fig.4 that the values obtained in the {\it "phonon"}
 approach (circles)
are systematically below the {\it "bremss"} evaluation, as expected
from the previous discussion. In order to get the {\it "bremss"}
results we need much larger values of the parameter $n_{GDR}^{(0)}$,
corresponding to times well before statistical equilibrium, not
consistent with the phonon model picture.
\begin{figure}[htb]
\epsfysize=5.5cm
\centerline{\epsfbox{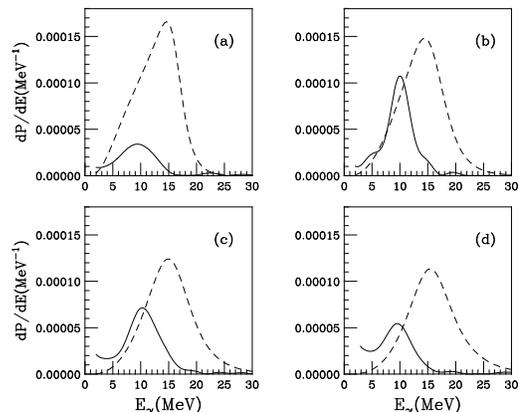}}
\caption{
Photon spectra for the $O$+$Mo$ system at the four energies as in Fig. 2:   
bremsstrahlung (solid line) and first step statistical
contribution (dashed line). 
}
\end{figure}

In Fig.4 we report also the ratio between total pre-equilibrium and
total statistical photon multiplicity
as a function of the initial excitation energy of the compoud system. 
This ratio is
a good measure of the pre-equilibrium effect which should be seen
in experiments when the total photon yields from a $N/Z$ asymmetric
reaction and a symmetric one, both forming the same $CN$ in similar
conditions (temperature, angular momentum) are compared.
To calculate the total statistical $GDR$ contribution we consider 
a very schematic
cascade Montecarlo model including only gamma and neutron channel
competition.
The optimum range of beam energies to observe the pre-equilibrium effect
appears to be around $8AMeV$. At higher beam energies a larger
statistical contribution to gamma decay will lower the relative
importance of pre-equilibrium contribution, which actually, 
as we have already discussed,
starts to decrease in absolute value. We have to remark that,
at the same $E^*$,  the effect is still
more important for the $Ca+Mo$ system, although now the
total statistical emission is also larger, see Eq.(\ref{statis}).
The $\simeq 16 \%$ enhancement is also in very good agreement
with the experimental estimate of \cite{fli96}.

\begin{figure}[htb]
\epsfysize=3.75cm
\centerline{\epsfbox{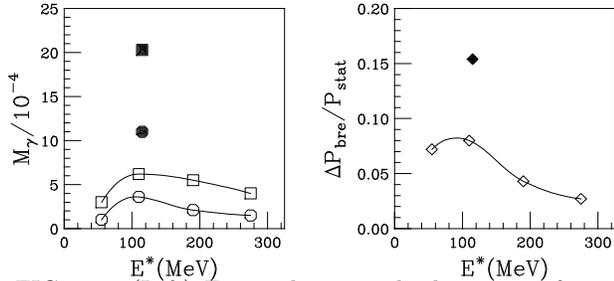}}
\caption{
(Left) Extra photon multiplicity as a function of initial excitation energy:
bremsstrahlung (empty squares), phonon model (empty circles)
for the $O$ + $Mo$ system  and $Ca$+$Mo$ (the corresponding full
symbols). (Right) Ratio of total pre-equilibrium to total statistical
$GDR$ $\gamma$-ray emission (see text).
}
\end{figure}

In conclusion, in this letter we have presented a bremsstrahlung 
approach to dipole
radiation from both thermal $GDR$ and dynamical dipole mode in 
fusion reactions. We stress that in our calculation we directly
get absolute values for the emission probabilities without
any normalization or adjusted parameters. Moreover the
{\it "bremss"} approach can be extended in a similar way
to dissipative deep-inelastic collisions \cite{pap99} where
similar prompt $\gamma-GDR$ emission has been recently
observed \cite{tro99}.

We have shown that when we consider the full dynamics of the
pre-equilibrium dipole in charge asymmetric fusion processes
the extra radiation contribution appears to be systematically
two times larger than the expectations, based on a
{\it phonon} model, used so far 
\cite{bar96,ced01,bar01,cho93,fli96,pie01} (see Fig.4).
The pre-equilibrium dipole radiation due to the charge asymmetry 
in the entrance channel can be then comparable to the
statistical $CN$ emission. In this sense it can represent
a new {\it "cooling"} mechanism in warm fusion
reactions to form less excited residues, e.g. of interest
for the synthesis of heavy elements.

A dependence of the pre-equilibrium gamma emission on beam energy has
been evidenced. We conclude that this is a consequence of the 
interplay between fusion
dynamics, damping properties of the dipole and statistical 
gamma emission. We expect a better observation of the effect in reactions
between nuclei with a larger initial dipole amplitude {\and} at beam 
energies where fusion dynamics is quite fast. Exotic (radioactive) beams
are a good opportunity for studying these phenomena. However a caution is
related to the fact that now a charge equilibration can take place also by 
an enhancement of fast nucleon emission.

We finally remark that the dynamical nature of the pre-equilibrium 
contribution will show up in a clear anisotropic $\gamma$-emission
since it is due to dipole oscillations on the reaction plane
(see details in \cite{bar01}). This could be a nice
characteristic signature of fusion paths in reactions induced
by radioactive beams.


\begin{references}
\bibitem{hof79} H. Hofmann et al., Z. Physik {\bf A 293} 229 (1979).
\bibitem{bon81} P. Bonche, N. Ngo, Phys. Lett. {\bf B 105} 17 (1981).
\bibitem{ditoro} M. Di Toro, C. Gregoire, Z. Physik {\bf A 320}  321 (1985).
\bibitem{sur89} E. Suraud et al., Nucl. Phys. {\bf A 492} 294 (1989). 
\bibitem{cho93} Ph. Chomaz et al. Nucl.Phys. {\bf A 563} 509 (1993).
\bibitem{bar96} V. Baran et al., Nucl. Phys. {\bf A 600} 111 (1996).
\bibitem{ced01} C. Simenel et al.  Phys.Rev.Lett. {\bf 67} 2971 (2001).
\bibitem{bar01} V. Baran et al., Nucl. Phys. {\bf A 679} 373 (2001).
\bibitem{brink90} D.M. Brink, Nucl. Phys. {\bf A 519} 3c (1990).
\bibitem{bor91} P.F. Bortignon et al.  Phys.Rev.Lett. {\bf 67} 3360 (1991).
\bibitem{fli96} S. Flibotte et al., Phys.Rev.Lett. {\bf 77} 1448 (1996). 
\bibitem{cin98} M. Cinausero et al., Nuovo Cimento {\bf A 111}  613 (1998).
\bibitem{amo198} F. Amorini et al., Phys.Rev. {\bf C 58} 987 (1998). 
\bibitem{pie01} D. Pierroutsakou et al., Nucl.Phys. {\bf A 683} 393c (2001).
\bibitem{bri55} D.M.Brink, D.Phil.Thesis, Oxford 1955.\\
          P.Axel, Phys.Rev. {\bf 126} 671 (1962).
\bibitem{jak62} J.D. Jackson, {\it Classical Electrodynamics} (Wiley,
New York, 1962).
\bibitem{ber98} A bremsstrahlung approach, T. Papenbrock and G.F. Bertsch 
Phys.Rev.Lett. {\bf 80}  4141 (1998) was applied to calculate the photon
emission during the alpha decay process. In our calculations the
classical {\it "bremss"} appears justified since the dynamics is well above
the threshold and so in a classical allowed region.
\bibitem{cal51} H.B. Callen, T.A. Welton, Phys.Rev {\bf 83}  34 (1951).
\bibitem{landau} L. D. Landau and E.M. Lifshitz, {\it
Statistical Physics}, (Pergamon Press, 1989), p. 386.  
\bibitem{sno86} K.A.Snover, Annu.Rev.Nucl.Part.Sci. {\bf 36}  545 (1986).
\bibitem{brink88} D.M. Brink, Nucl. Phys. {\bf A 482} 3c (1988).
\bibitem{pap99} M. Papa et al. , Eur.Phys.J. {\bf A 4}  69 (1999).
\bibitem{tro99} M. Trotta et al., RIKEN Review {\bf 23} 96 (1999),\\
~~~         M.Sandoli et al., Eur.Phys.J. {\bf A 6} 275 (1999). 

\end{references}
\end{document}